# Astronomical Research with Liquid Mirror Telescopes

Ermanno F. Borra,
Département de Physique, Université Laval, Canada G1K 7P4
Email: borra@phy.ulaval.ca
LMT WEB PAGE: http://astrosun.phy.ulaval.ca/lmt/lmt-home.html

Subject headings: Telescopes, Instrumentation, Cosmology

**ABSTRACT**

Liquid mirror telescopes are of interest to Astronomy because of their very low capital and operating costs. Low cost has a potentially revolutionary impact since it allows one to dedicate a large telescope to a narrowly focused project that needs a large quantity of observations. The technology is now well proven in the laboratory as well as in observatory settings and is stepping into the mainstream of astronomical research. This paper has two main purposes. a) To acquaint the astronomical community with LMT zenith observing and to illustrate the LMT advantage with specific examples of research topics of current interest that could be carried out only with LMTs because of their low costs. I argue that LMTs are more versatile than generally thought and that the challenge is to use LMTs in such ways as to maximize their advantages and minimize their disadvantages. I briefly review specific issues regarding zenith telescopes. I then illustrate the LMT advantage with specific examples of astronomical research. b) I show that the large number of objects observed with a surveying LMT allows one to carry out cosmological projects at low redshifts. I show that a variability survey would obtain light curves for several thousands of type Ia supernovae per year up to z=1 and easily discriminate among competing cosmological models. The same survey would find large numbers of MACHOs in clusters of galaxies and in background faint galaxies. Finally, I discuss a spectrophotometric survey carried out with interference filters, showing its power to discriminate among cosmological models and to study the large-scale distribution of galaxies in the Universe.

# 1. INTRODUCTION

Surveys have always played important roles in Astronomy but they are very demanding of telescope time, making it difficult to carry them out with large telescopes. Expensive large telescopes can only be justified by sharing them among large numbers of users. As a consequence, surveys are usually conducted with small telescopes. On the other hand, a number of important topics in contemporary Astronomy demand large numbers of observations of faint objects that can only be gathered by survey-like observing with large telescopes. This need has led to the emergence of the "key project" concept whereupon large groups of scientists team up to secure batches of telescope time. An example of this kind of observing is currently taking place with supernovae searches that endeavor to measure the main cosmological parameters (e.g. Filippenko & Riess 2000). However even key projects can only obtain relatively small numbers of nights per year. As a consequence, it takes years to gather relatively small amounts of data to reach conclusions that suffer from small number statistics as well as poorly understood systematic effects. These facts have been recognized by the astronomical community and a handful of dedicated survey telescopes in the 2.5-m to 4-m class are presently being built. Even larger survey telescopes are being planned (Tyson, Wittman & Angel 2000). However, these survey telescopes will be few in number for they are expensive: They will therefore be dedicated to general purpose surveys with some time dedicated to specialized short surveys and none to specialized time-demanding surveys.

In recent times, an inexpensive technology has emerged that uses liquid mirrors. Optical shop tests of liquid mirrors have demonstrated diffraction-limited performance (Borra et al. 1992, Girard & Borra 1997, Ninane & Jamar 1996). A 3.7-meter LM has been tested under severe perturbations in the laboratory, demonstrating robust performance (Tremblay & Borra, 2000). Several first generation liquid mirror telescopes in the 3-m class have been built and operated for a few years at the University of Western Ontario (Sica & Russel 1999), UCLA (Wuerker 1997) and at the NODO orbital debris observatory run by NASA ( Hickson & Mulrooney 1997)  demonstrating that the technology, although in its infancy, is sufficiently robust to be used for extended periods of time in observatory settings. The 3-m NODO LMT is extensively discussed by Mulrooney (2000) who describes its operation and performance. (see also [http://www.sunspot.noao.edu/Nodo/nodo.html](http://www.sunspot.noao.edu/Nodo/nodo.html) ). While the UWO and UCLA LMTs are used for the atmospheric sciences and are not imagers (but operate in very harsh climates), the NODO telescope is of particular importance to astronomy since it is an imager that has continuously operated for several years and is beginning to yield published astronomical research (Hickson & Mulrooney 1997, Cabanac & Borra 1998). Figure 1  shows a sample of the kind of data obtained by the NODO LMT. It compares the same region of sky observed with the NODO LMT and the Palomar 48-inch Schmidt. This first generation LMT will soon be followed by second generation LMTs: the 6-m LZT (Hickson et al. 1998) that will carry out spectrophotometry and the 4-m ILMT project that will carry out a wide-band imaging survey. LMTs are thus coming of age.

Liquid mirrors are of interest to Astronomy because of their very low costs (one to two orders of magnitudes less than a conventional glass mirror and its cell). Experience with the current LMTs shows that their operating costs are also very low. The "hassle factor" is low when compared to classical glass mirrors, for a glass mirror must be periodically cleaned, aluminized and adjusted in its cell, a cumbersome time-consuming process; by comparison, maintenance of an LM is virtually nonexistent, consisting of a monthly cleaning that lasts about an hour for a 4-m mirror. Low cost has a potentially revolutionary impact since it allows one to carry out narrowly focused projects that need dedicated large telescopes. On the other hand LMTs have a major limitation: They can only point at the zenith so that they only can observe inside a band of sky limited by the field accessible with the optical corrector of the telescope. Note, however, that technological developments may change this in the future with off-axis correctors and tiltable LMs (Moretto & Borra 1997, Borra, Ritcey &Artigau 1999). At present, the challenge to the observing community is: given an appropriate research topic, adopt an observing strategy that minimizes the limitations of LMTs while maximizing their advantages. Certainly, any project that can be done with a LMT can be done just as well, if not better, with a conventional tiltable telescope. However, in a real world of limited budgets, some projects that are utopical with conventional telescopes become quite feasible with inexpensive LMTs: This is where LMTs become interesting. The promise of a dedicated 4-m to 10-m diameter telescope working full time on ones pet project is sufficiently of an alluring prospect to make it worthwhile to contemplate its use.

The purpose of this paper is to acquaint the astronomical community with LMT zenith observing and to illustrate the LMT advantage with specific examples of research topics of current interest that could be carried out with LMTs. This is meant to illustrate the usefulness of LMTs to do frontier research; consequently, the topics addressed are purposely kept simple and of current interest. I have conservatively limited the discussion to 4-m and 6-m LMTs, since they are presently being built, the technology is well-proven in the laboratory (Tremblay & Borra 2000) and those sizes are a small extrapolation from the NODO 3-m, which has, at the time of this writing, operated for several years (Hickson & Mulrooney 1997). There is little doubt that LMTs in the 8-m class can be built using existing technology.

## 2. OBSERVING WITH A ZENITH TELESCOPE

As most astronomers are not familiar with zenith observing, it is useful to summarize here some of its characteristics. The present generation of transit telescopes track by using the Time Delayed Integration (TDI) technique (also known as driftscanning) with CCD detectors. TDI is implemented by drifting the potential wells that define the pixels of the CCD, in a readout mode, at the same speed as the image of the sky moves in the focal plane of the telescope. A LMT equipped with a conventional corrector would continuously monitor, with a CCD, a strip of sky passing through the zenith with a nightly integration time given by the time it takes an object to cross the CCD detector. Coadding successive nightly observations with a computer obviously can yield arbitrarily long integration times. On the one hand, while TDI zenith observing has limitations, on the other, it has advantages with respect to tracking observations. Flatfielding and defringing are much more accurate because the images are formed by averaging over entire CCD columns (in the direction of the scan). Extinction and seeing are optimized at the zenith. Last but not least, it yields 100% observing efficiency because there is no overhead from slewing, reading out the CCD (it is readout continuously), taking flatfielding frames, etc. There is a significant efficiency problem with large tracking telescopes, for reading-out a large format CCD is time consuming and CCDs saturate in a few minutes with a 4-m telescope observing with broadband filters.

**2.1 Integration times and limiting magnitude**s

TDI tracking with a zenith telescope restricts one to integration times limited by the time it takes an object to drift across the CCD detector. The nightly single-pass integration time with a zenith telescope is given by

$$t = 1.37 \times 10^{-2} \, n \, w / (f \cos(lat)), \quad (1)$$

where the time is in seconds, n is the number of pixels along the read-out direction of the CCD, w the pixel width (microns), f the focal length of the telescope (meters) and *lat* is the latitude of the observatory. It turns out that for telescopes and detectors of practical present interest, Equation 1 predicts integration times of the order of 100 seconds. Even such a short integration time yields surprisingly faint limits with a large LMT and efficient detector. Note that TDI tracking with a zenith telescopes aberrates the PSF but properly designed correctors (Hickson & Richardson 1998) can remove most of the aberration.

Let us consider a 4-m diameter f/1.95 LMT equipped with a 4096X4096 mosaic of 2048X2048 CCDs having 15-micron pixels ( 0.4 arcsec/pixel): It yields an integration time of 127 seconds/pass. I have estimated in Table 1 the limiting magnitudes in B, R and I using the IRAF task CCDTIME available on the NOAO web site, assuming the CCD prime focus camera at the Kitt Peak 4-m telescope. I have assumed zenith observations with 1 arcsecond seeing and a 7 day old moon. To estimate its performance in one observing season, I have further assumed 60 nights of observations per season, excluding thus moonlit time and including only the cloudless (for a good site) nights during which an object would actually cross the zenith. The limiting magnitudes would increase by 2.2. Table 1 also gives the performance after 4 observing seasons. Because observations in I can be carried out during moonlit nights, I have assumed 120 nights/year in I.

**2.2 Regions of the sky observed and object counts**

The telescope scans a strip of constant declination equal to the latitude of the observatory. Figure 2 converts equatorial into galactic coordinates and can be used to determine the regions of sky sampled by a zenith telescope in a given site. As the Earth rotates and the seasons change, the telescope scans a strip of constant declination moving in and out of the galactic plane. It must be noted that at the latitudes of the best terrestrial sites (+30, - 30 and + 20 degrees) the telescope samples interesting regions of the sky. At +30 degrees, it goes through the north galactic pole. At - 30 degrees, it goes through the south galactic pole and also into the bulge and the center of our galaxy. Four strips of sky observed by zenith telescopes located at selected terrestrial latitudes of interest are highlighted in the figure. The width of the strip of sky observed by the CCD is given by

$$S = 0.206 \; n \; w \; /f, \qquad (2)$$

where S is expressed in arcseconds, w is the pixel width in microns, and n is the number of pixels in the direction perpendicular to the scan. With a 4096X4096 CCD mosaic having 0.4 arcseconds pixels, the strip of sky is 27 arcminutes wide. Table 2 gives the areas of the strips of sky covered by a corrector correcting a 1 square degree field as well as for 4096X4096 CCD mosaics having 0.4 and 0.6 arcseconds pixels.

We can estimate the number of objects that one can observe in a strip of sky from the known number counts /square degree of quasars (from Hartwick & Schade 1990) and galaxies (from Lilly, Cowie, & Gardner 1991 and Metcalfe, Shanks, Fong, & Jones 1991), brighter than a given B magnitude, at the galactic poles. Table 3 gives the integrated counts a B<22 , B<24 and B<27.  Let us only consider the strip of "extragalactic" sky having galactic latitude > 30 degrees and assume that the optical corrector of the telescope yields good images over a 1 degree field, well within the performance of existing corrector designs. We could observe 100 to 200 square degrees of "extragalactic sky", depending on the latitude of the observatory , and, at a latitude of 30 degrees, roughly $3 \times 10^6$ galaxies and  50,000 quasars with B< 24. We would observe over 100 million galaxies at B<27.

### 3. COSMOLOGY WITH LMTs

Liquid mirror telescopes promise major advances for deep surveys of the sky and, in particular, cosmological studies. Cosmological objects are faint and, furthermore, cosmological studies tend to be statistical in nature, hence need large numbers of objects; one therefore needs considerable observing time on large telescopes. It is difficult to dedicate a large number of observing nights to narrowly defined projects with a conventional large telescope. On the other hand, an inexpensive LMT can be dedicated to a specific project. The outstanding limitation of LMTs, that they can only observe near the zenith, is not a significant handicap for cosmological surveys since the current prejudice is that the Universe is isotropic on a sufficiently large scale. Finding otherwise would be a major discovery.

There is a large variety of cosmological observations that can be done with the kind of data just discussed: It is, basically, the same sort of science traditionally carried out with Schmidt telescopes. Table 4 gives a short list, certainly not inclusive, of topics that could be addressed with a LMT. In the next sections we will consider two specific examples of survey science doable with LMTs: A wide band imaging survey and a low-resolution spectroscopic survey.

### 3.1 Galaxies

To have an approximate view of how the observations would sample the Universe, I have computed the redshift distribution expected from a survey having lower limiting magnitude $m_0$ and upper limiting magnitude $m_1$ from the usual cosmological integral

$$\frac{dN}{dz} = d\Omega \int_{m0}^{m1} \Phi(M) \frac{dV}{dz} dm \quad , \qquad (3)$$

where $\Phi(M)$ is the differential luminosity function (per unit magnitude) of galaxies, $M(H_0, q_0, z, m)$ the absolute magnitude, m the apparent magnitude, $d\Omega$ the surface area element and $dV(H_0, q_0, z)$ the cosmological volume element , $H_0$ the present epoch Hubble constant, $q_O$ the deceleration parameter and z the redshift. The models have been computed with the mix of galaxy types and the K-corrections described in Shanks, Stevenson, Fong, & MacGillivray (1984). This is an oversimplification for the luminosity function varies with morphology and the Shechter function is an average over all Hubble types. Furthermore, the luminosity function evolves in time and it is far from trivial to take evolution into account. I have approximately taken evolution into account by simply using the parameters of the luminosity function, appropriate for the redshift ranges and magnitudes involved, from Lilly et al. (1995). The uncertainties brought by these assumptions are large but tolerable for our purpose since we are only interested in an estimate of the redshift space sampled, rather than detailed modeling. Figure 3 shows the redshift distributions expected for surveys reaching 22nd, 24th and 27th blue magnitudes. Hudon & Lilly (1996) have computed Redshift distributions in the R band. It must be noted (e.g. Hudon & Lilly, 1996) that models underestimate galaxy counts at low apparent magnitudes for galactic evolution introduces an excess of high-redshift galaxies. By the same token, one must realize that the Universe will be sampled to higher redshifts than indicated in Fig. 3.

**3.2 Clusters of galaxies**

We can estimate the number and redshift distributions of clusters of galaxies observed in the area of sky A of the survey from

$$N = n(>M) A \int_{0.1}^{z} \frac{dV}{dz} dz, \qquad (4)$$

where $n(>M)$ is the space density of clusters of galaxies having mass greater than $M$. We use the mass function of clusters of galaxies $n(>M)$ determined by Bahcall and Cen (1993). Table 5 gives the number of clusters observable in the degree-wide strip of extragalactic sky as function of z and richness class. We can see a significant number of clusters for which, for example, we could determine the distribution of mass via lensing of background galaxies, although we will loose part of the cluster for those at the edges of the field. Figure 3 shows that there will be a sufficiently large number of images of galaxies with z > z(cluster) to do the mapping.

**3.2 Cosmology at intermediate redshifts**

Figure 3 and Table 5 indicate that the observations sample relatively low redshifts. Traditionally the strategy has been to observe at as high redshifts as possible since cosmological effects increase strongly with redshift. This strategy makes sense considering how little telescope time one can obtain in one year on a conventional large telescope and the resulting small number statistics one has to deal with. However, this strategy suffers from several inconvenients. A major drawback comes from the fact that evolutionary effects increase strongly with redshift. This makes it extremely difficult to use high redshift objects for cosmological purposes, for evolutionary and cosmological effects are intertwined and extremely difficult to disentangle. A second drawback comes from the small numbers of objects observed. This not only gives high statistical errors but, more worrisome, makes it difficult to study systematic errors. In our case, observing at relatively low redshifts, cosmological effects are small but we have massive quantities of objects that give very small statistical errors and, furthermore, allow us to understand systematic errors. Furthermore, evolutionary effects are small so that corrections should be more reliable because of the smaller extrapolation from z=0. Another advantage of working at low z comes from the fact that objects are brighter on average and therefore easier to observe spectroscopically, if needed, with conventional telescopes. Finally, the high surface densities and relatively high brightness' of the objects make it practical to efficiently observe them with multiobject spectrographs. Sections 4 to 6 will give specific examples of the power of LMTs surveys to discriminate between cosmological models at intermediate redshifts.

## 4. A WIDE BAND SURVEY: SEARCHING FOR SUPERNOVAE

Table 1 shows that 4 years of observing with a 4-m LMT gets the survey down to 27th B,R,I magnitudes, comparable to the present deepest surveys, but in a much larger area of sky (Table 2). The deep images can be used to find targets for HST, NGST, the VLT and the new generation of 8-m telescopes. If the LMT is located at the same site as a large telescope it finds, as a bonus, the targets at the best place of celestial real estate:

the zenith. Because a LMT observes the same regions of sky night after night, variability "comes for free"; hence variability studies are a natural application for LMTs. There is a daily sampling with all objects observed every clear night whenever the sky is sufficiently dark; for faint objects observed in the visible (bluer than the R band), this restricts one to about two weeks of moonless nights a month; there is no moon restriction for observations redder than R. Because the topics are of current interest, we will consider the use of the data to observe type Ia supernovae for cosmological purposes and MACHOs.

**4.1 Type Ia Supernovae**

Type Ia supernovae (Sne Ia) are particularly interesting because they are bright and good standard candles (Branch and Tammann 1992). Table 6 gives magnitudes as function of redshift obtained from Schmidt et al. (1998).

Comparison of Tables 1 and 6 shows that a single nightly pass can detect a SN at maximum light at redshifts as high as z=0.8. As a matter of fact, considering the 1+z time dilation factor, 6 nights can be safely binned at z>0.5, going 1 magnitude deeper, and higher than z = 1, in the I band. Binning 6 nights in I is quite realistic if one considers that the I limiting magnitudes are essentially independent on the phase of the moon. Light decay time corrections to canonical peak luminosities only need observations to 15 days in the rest frame, at which time the luminosity has decreased by less than 2 magnitudes below the peak, allowing one to observe useful light curves to z = 1.0.

**4.2 Predicted discovery rates.**

Pain et al. (1996) has estimated the rate of Type Ia supernovae at z ~ 0.4. They predict 34.4 (+23.9, -16.2) events/year/square degree for magnitudes in the range 21.3 < R < 22.3, corresponding to 0.3<z<0.5. Table 2 shows that a survey with a 4KX4K CCD covers an extragalactic strip of sky ( bII > 30 degrees) having a surface of the order of 100 square degrees so that there should be about 3000 events per year in the extragalactic strip of sky. Approximately 2/3 of the objects would be unobservable because they would lie outside of the nightly strip monitored, giving an estimate of about 1000 Sne Ia per year with $0.3<z<0.5$. Table 7 estimates the number of Sne Ia in other redshift ranges by naively extrapolating with the differential volume element $dV(z, H_0, q_0)/dz$ . It assumes $H_0 = 75$, $q_0 = ½$ and neglect evolution, an assumption that underestimates counts at $z>0.5$.

Given these estimates and the uncertainties in the rates themselves, one should expect that the survey would discover and obtain light curves for a few thousands of Type Ia Supernovae per year. This large number of objects illustrates the power of LMT surveys, considering that the two main SN surveys carried out to date have found about 100 SNe, with $0.1< z <1$, from observations painstakingly gathered over several years. Few of them have decent light curves. After a few years of observing the LMT survey will have several thousand SNe Ia with complete light curves. Anybody who has tried to understand systematic effects in survey data (like this writer) knows all too well how the need to subdivide the data in subsamples rapidly reduces the number of objects at hand, losing any effect sought within the haze of small number statistics. One needs large samples of objects to obtain sufficient statistics and to fully understand the material at hand. Riess at al. (1998) concludes that their detection of a cosmological constant is not limited by statistical errors but by systematic ones. They follow with the usual statement that more data is needed to quantify lingering systematic uncertainties.

A bottleneck may arise from the need to do follow-up spectroscopy with a conventional telescope. On the other hand, detailed multicolor light curves minimize, if not totally eliminate, the need for spectroscopic observations of the supernovae themselves. The large number of observations will also minimize contamination of the sample from rare peculiar SNe. Redshifts of the host galaxies can be efficiently obtained from multiobject spectroscopy. In the most pessimistic scenario, should follow-up spectroscopy of the SNe be necessary, the use of an LMT to find and follow supernovae would save precious discovery time on oversubscribed conventional telescopes to be reallocated for spectroscopic follow-up.

**4.3 Cosmology with type Ia supernovae.**

Branch and Tammann (1992) have reviewed the use of type Ia supernovae as standard candles and consider several cosmological research topics. There is now intense interest in the determination of the cosmological parameters with Type Ia SNe. It took several years for the two major teams involved to gather data on about 100 SNe. The 4-m survey would get several thousand multicolor light curves in a few years, allowing one to study and understand characteristics of the sample such as extinction and light-curve dependent luminosity effects. Other interesting applications would come from the several thousands calibrated radial rods so gathered. For example, this would allow one to study the velocity field and peculiar motions, mapping the mass distribution of the Universe. Finally, the rates of supernovae, of all types, can be used to infer the rate of massive star formation as function of z.

As a specific example, let us consider the discriminatory power of the survey among cosmological models. Figure 4 gives the magnitude-redshift relations for three "flat" cosmological models with different combinations of $\Omega_M$ and $\Omega_\Lambda$. For clarity, I have subtracted from all curves the magnitude-redshift relation for the $\Omega_M = 0.3$, $\Omega_\Lambda = 0.0$ model. The small vertical bars at z=0.2, z=0.4, z= 0.6, z=0.8 and z = =1 give the estimated 2 standard deviations error bars. To compute the error bars, I have assumed a 0.4 magnitude standard deviation for an individual SN measurement, 1000 SN / year/(0.2 z bin) and that the magnitude errors are normally distributed, obtaining thus a standard deviation of 0.0125 magnitude for one year-worth of observations. We can see that the data easily discriminate among the models. If a systematic effect, such as dust or evolution, mimics a particular signature, it would have to mimic it very closely indeed. Podariu, Nugent, & Ratra (2000) have computed cosmological models with time-varying cosmological constants. Their figure 1 shows that most of the differences among models occur for $0.2<z<1.0$.

**4.4 Other uses of the data**

The survey would give an unprecedented sample of variable stars and extragalactic objects. Among several possible uses:

- Variability of QSOs: Most QSOs have variable luminosities (Hawkins 2000) so that variability can be used to find them. There should be about 50,000 QSOs with $B<24$ in a one-degree strip of sky. One would discover a substantial fraction of them and gather information on their variabilities. QSOs can be identified by their peculiar B-R colors so that variability combined with large B-R colors would allow us to identify them with a high success rate. The QSO candidates would have high surface densities and could thus be efficiently reobserved with multiobjects spectrographs.

- Detached double line binaries: Among all types of variable stars, detached double line binaries are particularly interesting since these objects give information on basic stellar parameters (e.g. masses and radii) and, furthermore, are surprisingly good primary distance indicators. They are rare, hence one needs to obtain well-determined light curves for a very large number of objects.

- Microlensed galactic objects: Predicting the expected yearly rates of galactic MACHOs requires detailed modeling and is beyond the scope of this work. We can however make very approximate estimates using predicted rates for bulge stars (Griest et al. 1991), leading to an estimated rate at $10^{-7}<G<10^{-6}$. An estimate of the number of stars one will observe in the R band can be obtained from the table of star number densities in the V band given by Zombeck (1990). To V= 21, the number of stars per square degree varies between 200,000 in the galactic plane to 60,000 at a galactic latitude of 20 degrees, a region of sky that can be monitored for half a year and an area of 50 square degrees. The survey would thus observe over 5 million stars in half a year. We would therefore expect from 0.5 to 5 events per year multiplied by an efficiency factor < 0.5. This rate appears low, however note that the predicted rate for the MACHO project observations of bulge stars was also low (<1.1/year) but a significantly higher rate has been found.

- Gravitational lenses. The telescope would monitor the light variations in gravitational lenses having a range of masses. This subject is covered in the next section.

- Last but not least: Serendipity. The history of Astronomy is replete with unexpected discoveries obtained from new types of surveys. Quasars, pulsars, gamma-ray bursters come immediately to ones mind. This would be the first time in the history of Astronomy that one would gather such massive variability information at faint magnitudes in such large regions of sky.

## 5. A WIDE BAND SURVEY SEARCHING FOR MACHOS

Gravitational lensing has become a very active field of research since the discovery of the doubly imaged quasar Q0957+561 AB (Walsh, Carswell, & Weyman 1979). There is now a very large literature on the subject and a growing number of applications of gravitational lensing (Refsdal & Surdej 1994). Reading the review by Refsdal & Surdej (1994) one realizes that Liquid Mirror Telescopes can be used for a variety of research projects in gravitational lensing (e.g. finding new lenses, monitoring brightness variations, etc...). Surdej & Claeskens (private communication) have estimated that a 4-m LMT having the parameters assumed in this article would detect 50 new macrolensed quasars having separations greater than 1 arcsecond under the most unfavorable cosmological parameters. Photometric monitoring of the lenses could be used for a number of important projects. This includes determination of the Hubble constant via time delays and studies of the MACHO population from microlensing in the individual components of the lens. Follow-up spectroscopy of the lenses could be used to study the structure of the nucleus of the Quasars.

Mapping the mass distribution in clusters of galaxies, and thus the dark mass distribution, from the distortion introduced by the cluster on the images of background galaxies, is another interesting cosmological application of lensing. The number of clusters mapped so far is small, given the challenge of identifying high redshift clusters and, especially, imaging the fields to very faint magnitudes. Table 5 gives the number of clusters observable in the 175 square degree strip of extragalactic sky as function of z and richness class. We can see a significant number of clusters for which we can determine the distribution of mass, although we will loose part of the cluster for those at the edges of the field. Figure 3 shows that there will be a sufficiently large number of images of galaxies with $z > z(cluster)$ to be able to do the mapping. Given the depth of the observations, and the angular extent covered by the survey the data is also well suited to measure image distortions introduced by large-scale structures.

Microlensing of background quasars could also be used to study the MACHO population in the clusters. We shall consider this in greater details in the following section.

**5.1 Microlensing by clusters of galaxies**

To estimate the number of microlensing events in the survey, one has first to determine the optical depth for such events. In what follow we shall use the simple theory and equations from Walker & Ireland (1995). We shall assume that the lensed quasars are all at z =2 and that the lensed clusters are at 0.1<z<1.0 . The mass and radii used by Walker & Ireland (1995) are for a richness class two cluster. I correct those numbers for cluster richness by scaling the surface densities and cluster radii from the masses in Bahcall & Chen (1993). I then compute the optical depth in a 30X30-arcminute field per cluster of a given richness class. I then multiply those optical depths, for each richness class, by the number of clusters/redshift interval in Table 5. Finally, I add all of the optical depths so obtained. This number , $<\tau>$ =0.44, gives the global optical depth of the survey. Considering the assumptions used, this number is obviously a rough estimate. The expected number of microlensed quasars is then given by

$$N = n_0 <\tau> \Omega \qquad (5)$$

Where $n_0$ is the number of quasars per unit solid angle and $\Omega$ is the solid angle observed, which corresponds to a 30X30 arcminute field, considering the way $<\tau>$ has been computed, and not the total area of the survey. Conservatively assuming that we can monitor the light curves of quasars to B= 22, and a surface density of 200 QSO/square degree, one finds that there should be about 50 quasars in a 30X30 arcminutes field. Equation 5 thus predicts about 20 microlensing events in the clusters monitored by the survey. Walker & Ireland (1995) predict 0.8 events from monitoring a low-redshift cluster with the MACHO camera. Comparisons of these numbers illustrates the power of an LMT survey. In practice, because the time scales of the lensing events at the redshifts of interest will be greater than a month (as explained below), we will be able to use coadded observations and thus detect light variations to B> 22, accessing a larger surface density of quasars and thus predicting an even larger number of events.

Let us now consider the range of MACHO masses that the survey is sensitive to, and the time scales involved. This can be estimated from figure 1 in Walker (1999) who, incidentally, mentions that the mass limits quoted by Walker & Ireland (1995) are too pessimistic. With the usual caveats regarding the poorly known QSO source size, we find that, assuming several years of monitoring, a cluster redshift range such as ours (0.1<z<1) is sensitive to masses ranging between $10^{-4}$ and $10^{-2}$ solar masses. The time-scales involved vary between a month and a few years and are well matched to an LMT survey, which has a sampling time of one day and, presumably, a lifetime of several years. The distribution of optical depth with redshift and richness class shows a distribution with overweighing at low redshifts (a factor of 5 larger at z =0.15 with respect to z = 0.95) and small richness numbers (a factor of 10 larger at richness class 0 with respect to richness class 2). The data obtained therefore nicely complement lensing surveys done with conventional telescopes in low redshift clusters, which have short resolutions times, thus accessing lower mass ranges, but are limited to long monitoring times by the usual restrictions on telescope time and human resources.

5.2 Microlensing of quasars by foreground galaxies.

Distant quasars will also be lensed by MACHOs in foreground galaxies. To estimate the optical depth to microlensing, one can use an expression (Walker, 1999) valid for z<1,

$$<\tau> = \Omega_M z^2. \tag{6}$$

Assuming $\Omega_M=0.1$ and galaxies up to a redshift of 0.2, we obtain $<\tau> = 0.004$. Since a 30 arcminute-wide strip of sky would monitor of the order of 25,000 quasars, we expect of the order of 100 events caused by masses greater than $10^{-4}$ solar masses. The time scales involved (> 1 month) are well matched to the time resolution of an LMT.

**6. LOW RESOLUTION SPECTROPHOTOMETRY**

Let us consider a LMT carrying out an imaging survey through interference filters with a CCD tracking in the TDI mode. Such a survey is planned with the 6-m Large Zenith Telescope (LZT) under construction near Vancouver, Canada (Hickson et al. 1998).

6.1 The data

The data consists of images from which one can obtain morphologies, positions, magnitudes and spectrophotometry having the wavelength resolution of the filters (from 150 Å to 300 Å). The spectrophotometry has somewhat lower resolution than one is accustomed to with astronomical spectrographs but it is comparable to the resolution of the Oke Palomar multichannel photometer (Oke 1969). This has been one of the most productive astronomical instruments ever, producing some 300 articles (Oke, private communication): Imagine similar data for every object in 300 square degrees of sky. The NODO 3-m LMT has collected this type of data for several years ( Hickson & Mulrooney 1997) as part of a cosmological survey. A subset of the data has been used by Cabanac and Borra (1998) to search for peculiar objects.

**6.2 Redshifts from low resolution spectrophotometry.**

Redshifts give crucial cosmological information and one may wonder how accurately they can be determined from low-resolution spectrophotometry. They can be obtained from cross-correlation techniques that also classify the galaxy by its spectral type. Cabanac and Borra (1995) as well as Hickson, Gibson and Callaghan (1994) have carried out simulations to determine redshifts with the cross-correlation applied to noise-degraded spectral energy distributions (SEDs) of galaxies. Cabanac and Borra (1995) also applied a break-finding algorithm that detects the 4,000 Å break in galaxies. They assume the interference filters of the LZT survey and obtain, for a signal to noise ratio (S/N) of 10, an rms redshift error less than 6,000 Km/sec for the mix of all galaxy types, and rms morphological errors less than 0.14. The precision is improved to 3,000 Km/sec for early Hubble types.

The breakfinding algorithm offers some tantalizing results. Cabanac and Borra (1995) find that at S/N = 10, the software correctly assigns redshifts for 95% of E and S0 galaxies within 3,000 Km/sec decreasing to 75% for Sab and to 40% for Sbc. The success rates increase with S/N. They find that a major source of uncertainty arises because the software has a hard time distinguishing, in the presence of noise, the redshifted 4,000Å break from other spectral features. If a reliable algorithm can be found to correctly identify the 4,000 Å break among other spectral features, the precision will increase considerably. For example, at a S/N = 10 the breakfinder yields a standard deviation of 1,000 Km/sec for galaxies as late as Sb, if it correctly identifies the redshifted 4,000 Å break.

In conclusion, Montecarlo simulations indicate that morphological types can be obtained with a reasonable precision. For a S/N of at least 10, redshifts can be obtained with, at worse, a rms error of 6,000 Km/sec that can improve to 1,000 km/sec for early type galaxies, a sufficient accuracy for many cosmological studies, in particular studies of the large scale structure of the Universe. One may be skeptical of this kind of redshift accuracy quoted for low resolution and signal to noise ratio spectra; however Beauchemin and Borra (1994) successfully detected redshift peaks corresponding to large scale structure found independently by others. This was accomplished with a somewhat higher resolution (75Å) but with noisy photographic slitless spectra.

### 5.3 Performance of a 6-meter diameter LMT

With a 4096KX4096K 15 micron pixels CCD mosaic, one gets a 120 seconds integration time for a single TDI nightly pass for a site at a latitude of 32 degrees. The telescope will cover the wavelength region from 4,000 Å to 10,000 Å with 40 interference filters having logarithmically increasing widths and adequate overlap (Hickson et al 1998). As explained in Borra (1995), assuming a complete spectral coverage from 4000 Å to 10,000 Å, taking account losses to weather (in a good site) and technical problems, one would get 3 passes/filter/year for a total integration time of 360 seconds/filter. In 4 years we would get 12 passes for a total integration time of 1,440 seconds/filter. Table 8 gives the performance of the LZT observing through a 200 Å filter centered at 4400 Å. It shows that 4 years of observing would get us to B=24 with a S/N = 10, sufficient to get 6,000 Km/sec to 1,000 Km/sec redshift errors, and over B=24 with S/N =5, sufficient for rougher energy distributions and redshifts. The increase in sky brightness with wavelength is roughly compensated by the flux increase with wavelength for most faint galaxies, at least for $\lambda < 7,000$ Å.

The huge database given by nearly 200 square degrees of images could be used for a multitude of projects. Basically one could use the data for the same kind of research that can be done with a Schmidt telescope but with accurate magnitudes and variability or spectrophotometric information.

### 6.2 $q_0$ from galaxy counts

As an example of the impact of LMTs in cosmology, let us consider the determination of $q_0$. This is a notoriously difficult measurement since curvature is small at low redshifts and one therefore traditionally has tried to obtain it, with a variety of methods, from observations at high redshifts. A first difficulty arises because, having to observe far, one needs intrinsically bright (or large) objects that tend to be rare. A second, and worse, difficulty is caused by evolution effects, important given the great lookback time at high z, that have bedeviled efforts to get q0: geometry and evolution enter all tests and cannot be disentangled.

Volume tests give the most sensitive measurements for $q_0$. Consider for example, the number of objects /unit surface/unit redshift, it is given by

$$\frac{dN}{dz} = d\Omega N_0 c^3 [q_0 z + (q_0 - 1)(\sqrt{1+2q_0 z} - 1)]^2 / [H_0^3 (1+z)^3 q_0^4 \sqrt{1+2q_0 z}], \qquad (7)$$

where the symbols have the usual meaning and $N_0$ is the space density at z=0. As with all geometrical tests, the difference among the various geometries only becomes large for z >1, where only intrinsically bright rare objects are detectable (e.g. QSOs) and where evolution effects are large; but is small for z<0.3 where evolution is much smaller and where intrinsically fainter and more numerous objects (e.g. ordinary galaxies) are detectable. The difficulty is illustrated by the 3 curves in Figure 5 that show the counts predicted at z = 0.1, 0.3 and 0.5 as function of q0, normalized to the counts predicted for q0=0.5. Let us determine whether we live in an open or closed Universe and take the criterion that we can differentiate between q0 = 0.4 and q0 = 0.5 at the 5 $\sigma$ level. At z = 0.3, the ratio between the counts for q0 = 0.4 and for q0 = 0.5 is 1.05. If we ask for a 5 standard deviation discrimination and assuming that H0 and N0 are known, Poisson statistics demand a minimum of 10,000 objects. Poisson statistics cannot be blindly applied because galaxies are known to be clustered on a scale of a few Mpc; there will be an excess of variance in the cell counts compared to a random distribution. However, the correction factor that takes into account the departure from Poisson statistics scales as 1/volume so that the correction is negligible for the large volume of sky that we sample, provided galaxies remain clustered at z=0.3 as they are in the local universe. This can be checked with the data

Figure 3 shows that there are about 2000 objects/square degree with 0.25 <z<0.35 and B< 24. A telescope located at a latitude of 30 degrees observing a one-degree wide extragalactic strip of sky (bII<30º) containing 175 degrees . Allowing for galactic extinction, the telescope could observe about 400,000 galaxies with 0.25 <z<0.35 in the one-degree-wide strip accessible with a conventional corrector. It will not be possible to obtain redshifts for all objects and there will also be some loss due to some large nearby galaxies and the halos of bright stars. However, there will be plenty of ordinary elliptical and spirals to get several 10,000 redshifts. Note that over 1/3 of the galaxies should be ellipticals and early type spirals for which $\sigma$ ~ 1000 Km/sec can be obtained from the 4,000 Å break (Cabanac & Borra 1995).

In practice, one would have to obtain counts at various redshifts and use a least-squares fitting procedure and he would have to consider systematic effects that may cause spurious z-dependent gains or losses of objects (e.g. magnitude cutoff, redshift or photometric errors). The large quantity of data should leave us well-equipped to understand this. For example, there is evidence that the faint end of the luminosity function evolves at surprisingly low redshifts. The data would allow us to measure the evolution and either correct for it or simply truncate the luminosity function at the appropriate magnitude. Close attention shall have to be paid to effects peculiar to the data, such as the effect of the large redshift errors, which depend on the Hubble type. This is not the place to carry out a detailed discussion; the relevant point is that the data allows us to contemplate such a project at all. The theme of this paper is that having a large telescope dedicated to a project allows us to consider a project that would be otherwise be unthinkable.

**6.3 Large Scale structure**

Figure 3 shows that the universe is sampled to significantly higher redshifts than any other large scale redshift existing (e.g.Center for Astrophysics) or planned (e.g. the Sloan Digital Survey ) albeit with a considerably lower radial velocity precision. One could therefore study the large scale structure of the universe from a few times the radial velocity precision to the redshift depth of the survey (a few Gpc). This complements the information that will be obtained from the other more precise but shallower surveys. Figure 6 is adapted from Vogeley (1995). It compares 1 $\sigma$ uncertainties, expected for a volume limited (to M*) sample of the SDSS northern redshift survey and assuming Gaussian fluctuations, to power spectra for CDM with different $\Omega h$. Power bars on smaller scales are of similar or smaller size than the symbols. The shaded box shows the range of the HPBW of the z distribution of the LZT survey (see Fig. 3). The HPBW of the LZT survey extends well beyond the HPBW sampled by the SDSS. Because the total numbers of galaxies are similar in the 2 surveys, we can expect a similar distribution of error bars. The error bars in the shaded box will approximately have the sizes of the error bars of the SDSS for $\lambda<200$ MPc. We can see that the LZT survey samples, with small error bars, the very long wavelengths at which the differentiation among the theoretical models is the greatest, and overlaps with the scales probed by COBE. Hopefully, the small error bars given by the large statistics may be able to detect the features predicted by some models; if needed the statistics can be increased by observing different strips of sky. The telescope could be moved or additional ones build, an acceptable alternative given the low cost of the system.

Because we get energy distributions, morphologies and accurate photometry we can repeat the analysis as function of Hubble type, spectral type, etc.... We also can study the redshift dependencies of the energy distributions and mixes of Hubble types.

**6. CONCLUSION**

Liquid mirror telescopes have come of age for the technology is well-proven both in the laboratory, where it has demonstrated high optical qualities and robustness, and in observatory settings where it has demonstrated scientific results and long-term reliable performance. The low cost advantage of liquid mirror telescopes allows one to dedicate one to a narrowly defined project that would be unpractical with a classical telescopes: This has a potentially revolutionary impact.

This paper discusses some particularities of LMT zenith observing and illustrates the LMT advantage with specific examples of research topics of current interest that, in a World of limited budgets, can be carried out only with LMTs because of their low costs. It shows that the large number of objects observed with a surveying LMT allows one to carry out cosmological projects at relatively low redshifts. It shows that a variability survey would obtain light curves for several thousands of type Ia supernovae up to $z=1$ and easily discriminate among competing cosmological models. The same survey would find large numbers of MACHOs in clusters of galaxies and in background faint galaxies. Finally, It discusses a spectrophotmetric survey carried out with interference filters, showing its power to discriminate among cosmological models and to study the large scale distribution of galaxies in the Universe.

There is now a window of opportunity for "early-bird" users of the technology to gather "low-hanging" fruits. The challenge is to use LMTs in such ways as to maximize their advantages and minimize their disadvantages. The history of Astronomy shows that whenever a new frontier (e.g. radio waves, $\gamma$-rays) has been opened, Astronomy has made huge strides, often led by unexpected discoveries (quasars, γ-ray bursts). Liquid mirror telescopes open a new frontier: The statistical frontier of extremely large numbers.

REFERENCES

Bahcall, N. A, & Cen, R. 1993, ApJ 407, L49.

Beauchemin, M. , & Borra, E.F. 1994, MNRAS 270, 811

Borra, E.F. 1995, Can J. of Phys. 73, 109.

Borra, E.F., Content, R., Girard, L., Szapiel, S., Tremblay, L.M., & Boily, E. 1992 ApJ 393, 829

Borra, E.F., Ritcey, A.M., & Artigau, E. 1999, ApJ Letters 516, L115.

Branch, D., & Tammann, G.A. 1992, ARAA 30, 359.

Cabanac, R., & Borra, E.F. 1998, ApJ 509, 309.

Cabanac, R., & Borra, E.F. 1995, PASP 108, 271.

Filippenko, A.V., & Riess, A.G. 2000, preprint (astro-ph/0008057).

Girard, L., & Borra, E.F. 1997, Applied Optics 36, number 25, 6278.

Griest, K. et al. 1991, ApJ 372, 79.

Hartwick, F. D. A., & Schade, D. 1990, ARAA 28, 437.

Hawkins, M.R.S. 2000, A &A 143, 465.

Hickson, P., Borra, E. F., Cabanac, R., Chapman, S. C., de Lapparent, V., Mulrooney, M., Walker, G. A. 1998, Proc. SPIE Vol. 3352, p. 226-232, Advanced Technology Optical/IR Telescopes VI, Larry M. Stepp; Ed.

Hickson, P., Gibson, B.K., & Callaghan, A.S. 1994. MNRAS 267, 911.

Hickson, P., & Mulrooney, M. K. 1997, ApJ Suppl. 115, 35.

Hickson, P., & Richardson, E.H. 1998, PASP 110, 1081.

Hudon, J.D., & Lilly, S. J. 1996, ApJ 469, 519.

Lilly, S. J., Tresse, L., Hammer, F., Crampton, D., & Le Fevre, O. 1995, ApJ 455, 108.

Lilly, S.J, Cowie, L.L., & Gardner, J.P. 1991, ApJ 369, 79.

Metcalfe, N., Shanks, T., Fong, R., & Jones, R.L. 1991, MNRAS 249, 498.

Moretto, G., & Borra, E.F. , 1997, Applied Optics - Optical Technology, Vol. 36, No. 10, 2114.

Mulrooney, M.K. 2000, unpublished Ph.D. thesis.

Ninane, N.M., & Jamar, C.A. 1996, Applied Optics 35, No 31, 6131.

Oke, J.B. 1969, PASP 81, 478.

Pain, R. et al. 1996, ApJ 473, 356.

Podariu, S., Nugent, P., & Ratra, B. 2000, preprint (astro-ph/0008281).

Refsdal, S., & Surdej, J 1994, Rep.Prog. Phys. 56, 117.

Riess, A.G. et al. 1998, AJ 116, 1009.

Shanks, T., Stevenson, P. R. F., Fong, & R., MacGillivray, H. T. 1984, MNRAS 206, 767.

Sica, R.J., & Russell, T. 1999, J. Atmo. Sci. 56, 1308.

Schmidt, B.P., et al. 1998 ApJ 507, 46.

Tremblay, G., & Borra, E.F. 2000, Applied Optics, Vol. 39, No 36, 5651

Tyson, J. A., Wittman, D.M. & Angel, J. R. P. 2000, preprint (astro-ph0005381)

Vogeley, M.S. 1995, in Clustering in the Universe, Proc. Of the XXXth Moriond conference.

Walsh,, D., Carswell, R.F., & Weynman, R.J. 1979, Nature 279, 38.

Walker, M.A., & Ireland, P.M. 1995, MNRAS 275, L41.

Walker, M.A. 1999, MNRAS 306, 504.

Wuerker, R. 1997, Opt. Eng. 36, 1421.

Zombeck, M. V. 1990), Handbook of Astronomy & Atsrophysics, Cambridge University Press, Cambridge, UK.

Table 1

Broad-band limiting magnitudes.

(4-m LMT with a 4096x4096 CCD mosaic having 0.4-arcsecond pixels, S/N = 5)

===============================================================

| 1 night (127 sec) * | | | 1 season (7600 sec)* | | | 4 seasons (30,500 sec)* | | |
|---|---|---|---|---|---|---|---|---|
| B | R | I | B | R | I | B | R | I |
| 24.4 | 24.1 | 23.7 | 26.6 | 26.3 | 25.9 | 27.4 | 27.1 | 26.7 |

\* Because observations in I can be carried out during moonlit nights, I have assumed integration times twice as long in I.

Table 2

Areas of sky observed from a site at ± 30 ° latitude

==================================================================

(square degrees)

| Field | Total | ExtraGalactic* |
|---|---|---|
| 1 degree | 312 | 156 |
| 4096x0.62 arcsec | 220 | 110 |
| 4096x0.41 arcsec | 146 | 72 |

* Regions having galactic latitudes > 30 degrees.

Table 3

Integrated object counts

1 degree-wide strip of "extragalactic sky "(bII>30º)

============================================================

|  | B<22 | B<24 | B<27 |
|---|---|---|---|
| Galaxies | 200,000 | 3,000,000 | 200,000,000 |
| Quasars | 10,000 | 50,000 | ? |

Table 4

SELECTED COSMOLOGICAL TOPICS THAT CAN BE STUDIED WITH AN LMT

===============================================================

Preselection of objects for further studies with conventional telescopes

Evolution of the luminosity function of galaxies

Evolution of galaxies

Large scale structure of the universe (with redshifts and/or positions)

Supernova rates

Search for gravitational lenses

Gravitational potentials of clusters and large scale structures with arcs and arclets.

Variable objects (e.g. AGNs)

Fluctuations of faint galaxy counts to identify distant clusters.

Find QSOs and AGNs by variety of techniques (astrometry, variability, spectra, colors)

Search and study of primeval galaxies

Serendipity (This is arguably the most exciting one)

Weak lensing
______________________________________________________________________

Table 5

Integrated number counts from z=0.1 and z = $z_{final}$ of clusters of galaxies in the degree-wide strip of extragalactic sky as function of richness class (H0 = 75, $q_o$= 1/2)

==============================================================

| R, $z_{final}$ | 0.2 | 0.3 | 0.4 | 0.5 | 0.6 | 0.7 | 0.8 | 0.9 | 1.0 |
|---|---|---|---|---|---|---|---|---|---|
| 0 | 26 | 81 | 167 | 280 | 419 | 579 | 757 | 950 | 1154 |
| 1 | 11 | 36 | 74 | 124 | 186 | 257 | 336 | 422 | 512 |
| 2 | 2 | 7 | 15 | 25 | 37 | 51 | 67 | 84 | 102 |
| 3 | 0 | 1 | 2 | 3 | 5 | 6 | 8 | 10 | 13 |

Table 6

Apparent R and I magnitude of a type Ia supernova at maximum light as function of redshift

==================================================================

| z | $m_R$ | $m_I$ |
|---|---|---|
| 1.0 | 25 | 23.8 |
| 0.8 | 23.8 | 23.0 |
| 0.6 | 22.7 | 22.5 |
| 0.4 | 21.8 | 21.6 |
| 0.2 | 20.2 | 20.2 |

Table 7

Predicted Sne Ia discovery rates in $\Delta z = 0.2$ bins.

========================================================================

| z | events/year |
|---|---|
| 0.2 | 500 |
| 0.4 | 1000 |
| 0.6 | 1500 |
| 0.8 | 1900 |
| 1.0 | 2000 |

Table 8

Performance of the LZT in a narrow-band blue filter. *

=================================================================

|  | Narrow-band filters (200Å centered at 4400Å) | |
| --- | --- | --- |
| Duration of the survey | S/N=10 | S/N= 5 |
| 1 night, 1 pass (120 seconds) | 22.4 | 23.2 |
| 1 year, 3 passes (360 seconds) | 23.0 | 23.8 |
| 4 years, 12 passes (1,440 seconds) | 23.7 | 24.5 |

_________________________________________________________________

*Assumes a 4Kx4K CCD.

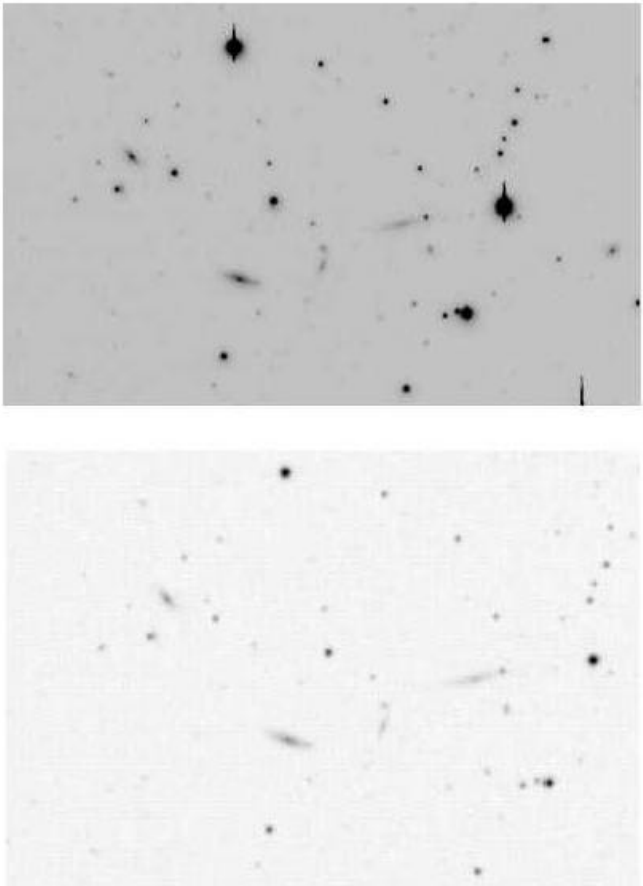

Figure 1: It shows a sample of the kind of data obtained by the NODO LMT. It compares the same region of sky observed with the NODO LMT (top) and the Palomar 48-inch Schmidt (bottom). The field is 5 arcminutes X 7 arcminutes and centered at 12h 08 m and + 33 degrees (2000.0 coordinates). Better (and additional figures) can be found in various LM web sites (http://wood.phy.ulaval.ca/home.html

http://www.astro.ubc.ca/LMT/lmt.html http://www.sunspot.noao.edu/Nodo/nodo.html

40Figure 2

Converts equatorial into galactic coordinates and can be used to determine the regions of sky sampled by a zenith telescope in a given site. As the Earth rotates and the seasons change, the telescope scans a strip of constant declination moving in and out of the galactic plane.

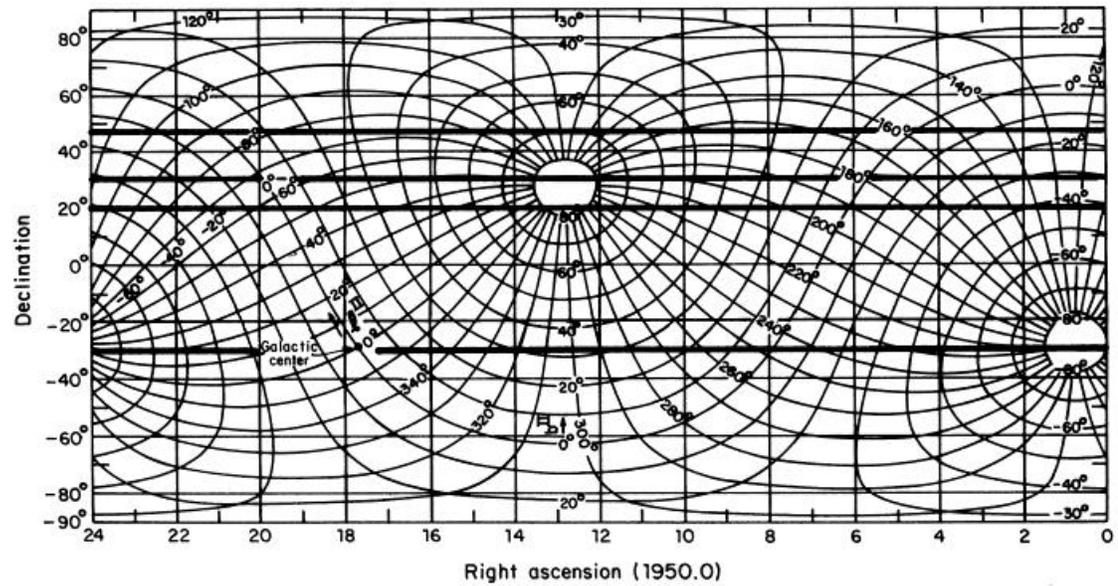

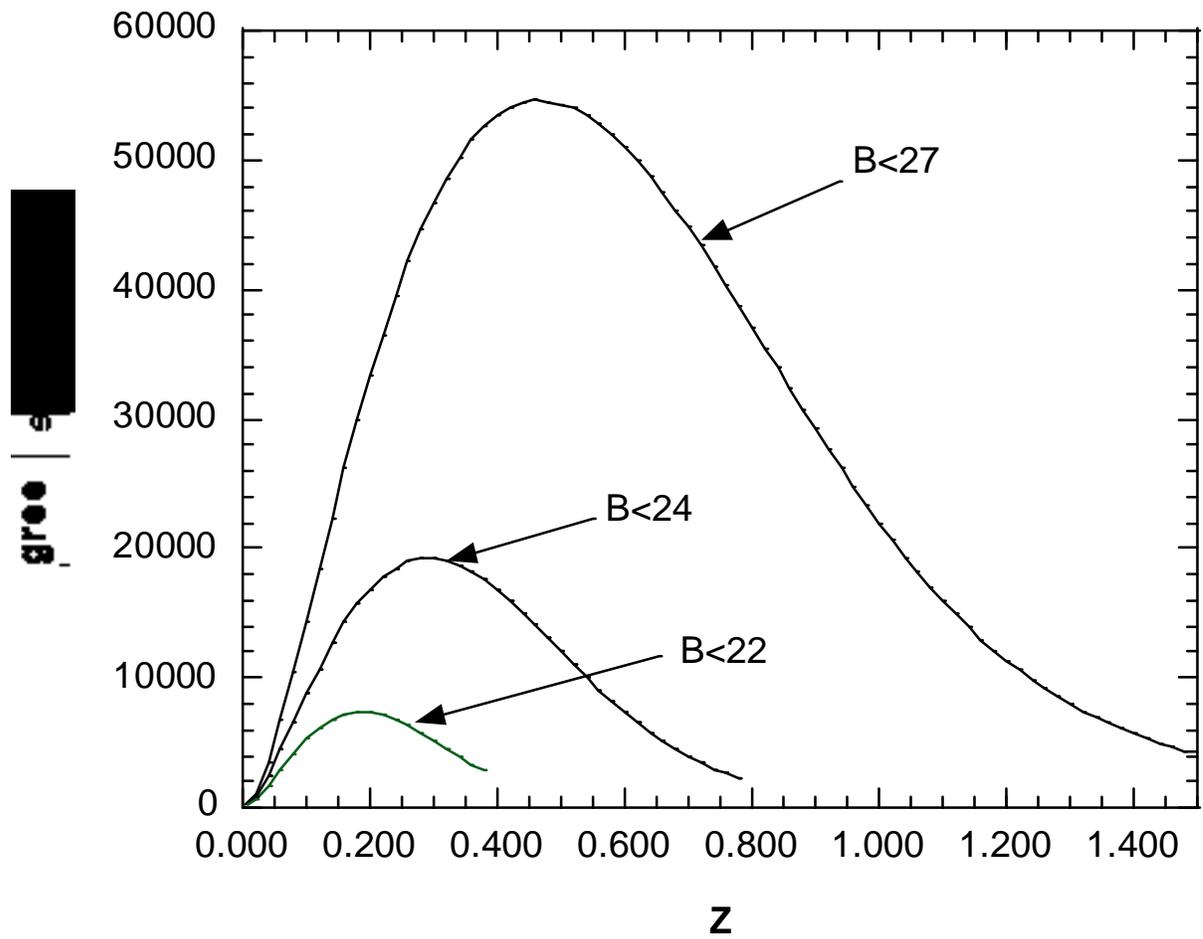

Figure 3

Redshift distributions expected for surveys reaching 22nd, 24th and 27th blue magnitudes.

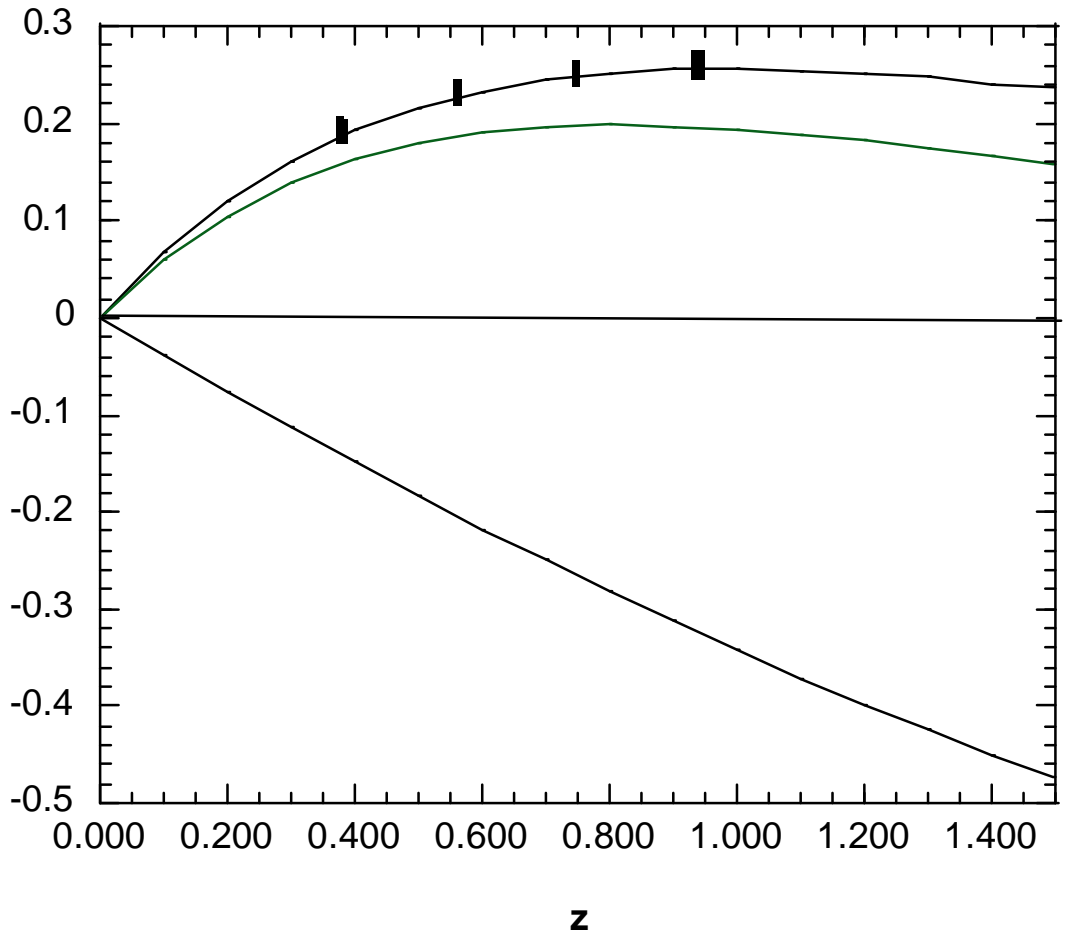

Figure 4

Gives the magnitude-redshift relations for three "flat" cosmological models with different combinations of $\Omega_M$ and $\Omega_\Lambda$ (top to bottom $\Omega_M = 0.3$, $\Omega_\Lambda = 0.7$; $\Omega_M = 0.35$, $\Omega_\Lambda = 0.65$, $\Omega_M = 1.0$, $\Omega_\Lambda = 0.0$). For clarity, the magnitude-redshift relation for the $\Omega_M = 0.3$, $\Omega_\Lambda = 0.0$ model has been subtracted from all curves. The small vertical bars give 2 standard deviations error bars estimated for the supernova survey discussed in section 4.

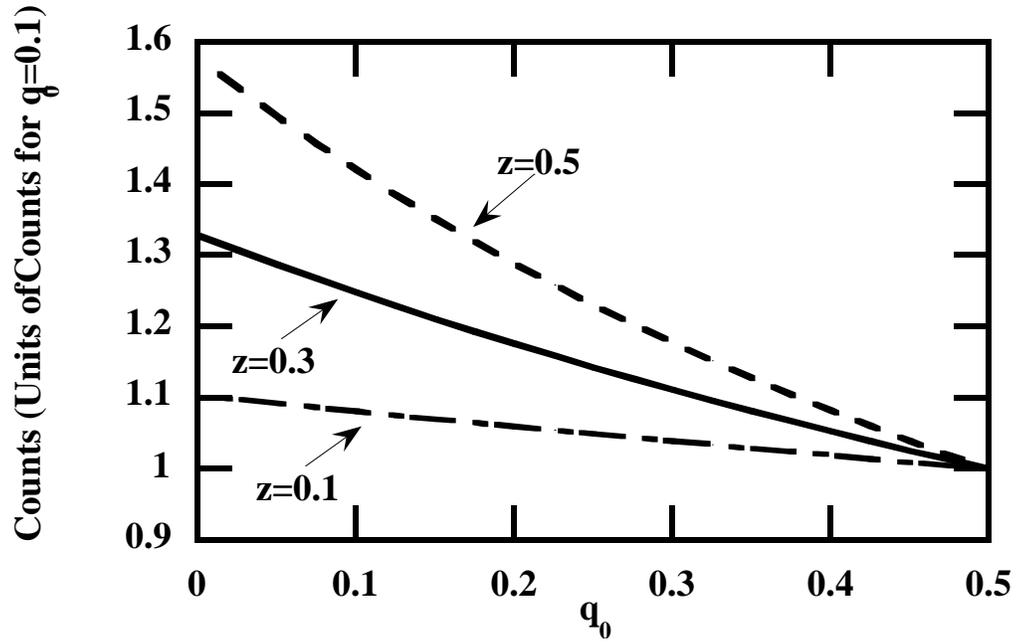

Figure 5

The 3 curves in figure show the galaxy counts predicted at z = 0.1, 0.3 and 0.5 as functions of $q_0$, normalized to the counts predicted for $q_0=0.5$.

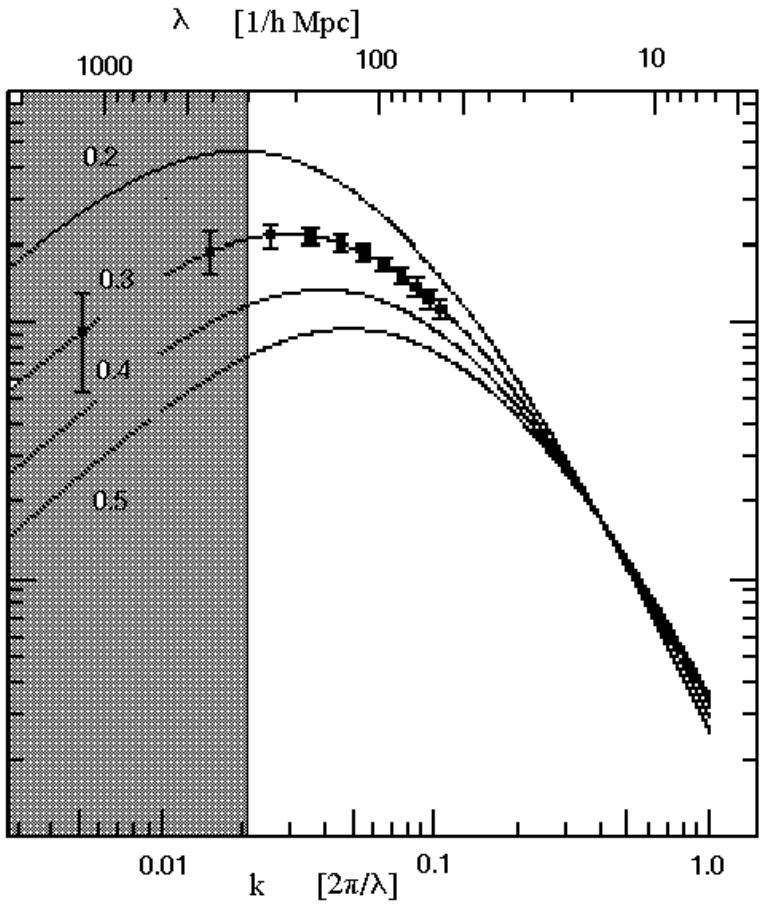

Figure 6

Adapted from Vogeley (1995), it compare 1 σ uncertainties, expected for a volume limited (to M*) sample of the SDSS northern redshift survey and assuming Gaussian fluctuations, to power spectra for CDM with different Ωh. Power bars on smaller scales are of similar or smaller size than the symbols. The shaded box shows the range of the HPBW of the z distribution of the LZT survey.